\documentclass[vecphys]{svmult}

\usepackage{graphicx}        % standard LaTeX graphics tool
\usepackage[bottom]{footmisc}% places footnotes at page bottom
\usepackage{amssymb,amsmath}
\newcommand{\tref}[1]{(\ref{#1})}

% ...
\newcommand{\tnote}[1]{} % private footnote, not even for preprint
\newcommand{\tpre}[1]{} % Comments for preprint version
\newcommand{\tprenote}[1]{} % Footnote only for preprint version

\renewcommand{\tpre}[1]{#1} % Comments for preprint version
\renewcommand{\tprenote}[1]{\footnote{#1}} % Comments for preprint version

% ...
\newcommand{\tsevec}[1]{\mathbf{#1}}
\newcommand{\tsemat}[1]{{\mathbf{\textsf{#1}}}}
\newcommand{\bea}{\begin{eqnarray}}
\newcommand{\eea}{\end{eqnarray}}
\newcommand{\beq}{\begin{equation}}
\newcommand{\eeq}{\end{equation}}
\newcommand{\nnel}{\nonumber \\ {}}

\newcommand{\ra}{\rightarrow}

\newcommand{\fvec}{\tsevec{f}}

\newcommand{\rhoexp}{\langle\rho\rangle}

\newcommand{\Tmat}{\tsemat{T}}

\newcommand{\tav}[1]{\langle #1 \rangle}
\newcommand{\kav}{\tav{k}}
\newcommand{\ksqav}{\tav{k^2}}

\newcommand{\efunc}{\omega}

\newcommand{\fMa}{f^{(MA)}}
\newcommand{\ftwoa}{f^{(2A)}}
\newcommand{\fzero}{f^{(0)}}

\newcommand{\GMa}{G^{(MA)}}

\newcommand{\lMa}{\lambda_{MA}}
\newcommand{\lzero}{\lambda_{0}}

\newcommand{\ltwoa}{\lambda_{2A}}

\begin{document}

\title*{Network Rewiring Models}

% Use \titlerunning{Short Title} for an abbreviated version of
% your contribution title if the original one is too long
\titlerunning{Network Rewiring Models}
% Use \authorrunning{Short Title} for an abbreviated version of
% your contribution title if the original one is too long

\author{T.S.\ Evans\inst{1}\and A.D.K.\ Plato\inst{2}}

\institute{Theoretical Physics,
 Blackett Laboratory, Imperial College London,
 South Kensington campus, London, SW7 2AZ,  U.K.
\and Institute for Mathematical Sciences, Imperial College London, 53 Prince's Gate
South Kensington, London, SW7 2PG, U.K.}
%
% Use the package "url.sty" to avoid
% problems with special characters
% used in your e-mail or web address
%

 \maketitle

\tpre{
\section*{Abstract\tprenote{30th April 2007,
\texttt{Imperial/TP/07/TSE/2}, \texttt{arXiv:0707.3783}.  This is a
longer version of contribution accepted for ECCS07.}} Recently we
showed that a simple model of network rewiring could be solved
exactly for any time and any parameter value. We also showed that
this model can be recast in terms of several well known models of
statistical physics such as Urn model and the Voter model. We also
noted that it has been applied to a wide range of problems. Here we
consider various generalisations of this model and include some new
exact results. }

% *******************************************************
\section{Introduction}\label{sintro}

Graphs with a constant number of edges and vertices but which evolve
by rewiring those edges are a classic network model as exemplified
by Watts and Stogatz \cite{WS98} (\cite{PLY05,OTH05} provide further
examples). Such network evolution may also be recast as other types
of statistical physics models (for example see
\cite{GL02,EH05,Liggett99,SR05}). As many real systems are
effectively of constant size, non growing networks can also be used
to model a wide range of data: the transmission of cultural
artifacts such as pottery designs, dog breed and baby name
popularity (as in \cite{HB03,HBH04,BS05,BLHH07}),  the distribution
of family names in constant populations, and the diversity of genes.
In this paper we look at various extensions to a model of network
rewiring for which an exact solution \cite{Evans07} was presented at
ECCS06 \cite{EP07a} and in more detail in \cite{EP07b}.

% *******************************************************
\section{The Model}\label{smodel}

\begin{figure}[hbt]
\centering
\includegraphics[width=10cm]{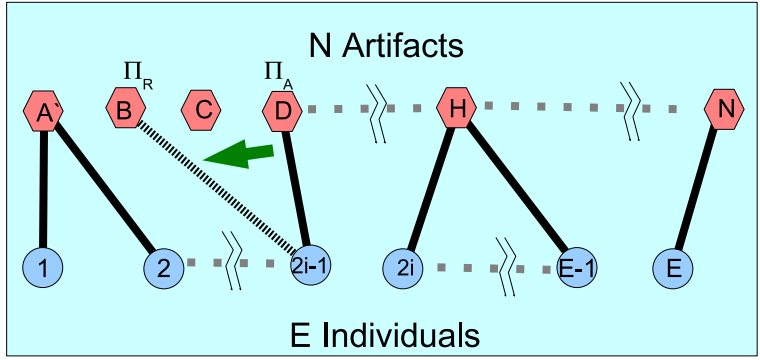}
\caption{The bipartite graph has $E$ `individual' vertices, each
with one edge. The other end of the edge is connected to one of $N$
`artifact' vertices. If the degree of an artifact vertex is $k$ then
this artifact has been `chosen' by $k$ distinct individuals. At each
time step a single rewiring of the artifact end of one edge occurs.
An individual is chosen (number $(2i-1)$ here) with probability
$\Pi_R$ which gives us the departure artifact (here D).  At the same
time the arrival artifact is chosen with probability $\Pi_A$ (here
labelled B). After both choices have been made the rewiring is
performed (here individual $(2i-1)$ switches its edge from artifact
D to B).}
 \label{fCopyModel4}
\end{figure}

We will study the rewiring of a bipartite graph consisting of $E$
`individual' vertices connected by one edge only to any one of $N$
`artifact' vertices, as shown in Fig.~\ref{fCopyModel4}.
At each time step two choices are made.  With probability $\Pi_R$ an
individual is chosen.  It is the artifact end of its single edge,
connected to the `departure' artifact, which is to be rewired. An
`arrival' artifact is also selected with probability $\Pi_A$. Only
after the choices are made is the network altered by rewiring the
chosen edge so that its artifact end is moved from the departure to
the arrival artifact.  Note we do not explicitly exclude the
possibility that the departure and arrival artifacts are the same.
The individual vertices always retain one edge while the degree $k$
of the artifact vertices is changing in time, only its average
degree $\kav = E/N$ is constant. It is the distribution of the
artifact vertices at time $t$, $n(k,t)$, and its probability
distribution $p(k,t)=n(k,t)/N$, that we study. This process can be
viewed in many other ways \cite{EP07b}.

The evolution of the degree distribution in the mean field
approximation is described by the master equation
\cite{Evans07,EP07a,EP07b}
\begin{eqnarray}
\lefteqn{n(k,t+1) - n(k,t)}
 \nnel
 &=&   n(k+1,t) \Pi_R(k+1,t) \left( 1- \Pi_A(k+1,t) \right)
 \nnel
 &&
     - n(k,t)   \Pi_R(k,t)   \left( 1- \Pi_A(k,t)   \right)
     - n(k,t)   \Pi_A(k,t)   \left( 1- \Pi_R(k,t)   \right)
 \nnel
 &&
     + n(k-1,t) \Pi_A(k-1,t) \left( 1- \Pi_R(k-1,t) \right)
 \, .
   \label{neqngen}
\end{eqnarray}
For our physical problem the removal probability must always satisfy
$\Pi_R(k=0)=0$ and $\Pi_R(k=E)=1$. In addition for physical
solutions we must have $n(k,t)=0$ if $k<0$ or $k>E$. The presence of
the factors of $(1-\Pi)$ ensure that if the degree distribution
initially satisfies its physical boundary condition, $n(k,t=0)=0$ if
$k<0$ or $k>E$, then this boundary condition is automatically
satisfied at all times\tprenote{For this to be true it is absolutely
vital that we have the factors of $(1-\Pi_R(k))$ to ensure that with
the condition $\Pi_R(k=E)=1$ we do not include processes where an
artifact with $E$ edges is lost because we are adding another edge
(the third term in \tref{neqngen} for $k=E$).}.  The factors of
$(1-\Pi)$ are not seen in the master equations of the literature
\cite{PLY05,OTH05,GL02,DM03} and correspond to events where the
arrival and departure artifacts are chosen to be the
same\footnote{\tpre{These events occur with probability
$(\Pi_R\Pi_A)$. Since the network is unchanged by such events, we
must exclude such events from the evolution of $n(k,t)$ and the
factors of $(1-\Pi)$ implement this.} It is an approximation to drop
these terms.  In our model this is not justified for certain
parameter values.}.

In general the master equation \tref{neqngen} gives the evolution
only in the mean-field approximation because we are taking an
ensemble average over many instances of the stochastic evolution and
using the product of averages where we should have the average of
products. However when the normalisations of probabilities $\Pi_A$
and $\Pi_R$ are constant then the master equation \tref{neqngen} may
be exact.  The most general $\Pi_R$ and $\Pi_A$ for which this is
true is
\beq
 \Pi_R = \frac{k}{E}, \qquad
 \Pi_A = p_r\frac{1}{N} + (1-p_r)\frac{k}{E},
   \qquad (E \geq k \geq 0) \; .
 \label{PiRPiAsimple}
\eeq
We will restrict ourselves to these forms and therefore our analytic
results are \emph{exact} \cite{EP07b}.  Thus we are choosing our
arrival edge with a mixture of preferential attachment (probability
$(1-p_r)$) and random attachment (probability $p_r$). The removal
artifact is found by choosing the artifact end of a randomly
selected  edge --- `preferential removal'.  The use of probabilities
proportional to $k$ can emerge naturally through short range
searches of many networks, since the probability of arriving at a
vertex on a random graph is proportional to its degree
\cite{NSW01,DMS03a,FFH05}.

Not only is the master equation exact for our chosen probabilities
\tref{PiRPiAsimple} but the exact solution for the degree
distribution $n(k,t)$ may be found for any finite parameter value.
This may be done in terms of $(E+1)$ eigenfunctions
$\efunc^{(m)}(k)$ and their corresponding generating functions
$G^{(m)}(x)$
\bea
 n(k,t) = Np(k,t) &=& \sum_{m=0}^{E} c_m (\lambda_m)^t
 \efunc^{(m)}(k) \, ,
 \\
 G(x,t) & := & \sum_{k=0}^{E} x^k n(k,t)
  = \sum_{m=0}^{E} c_m (\lambda_m)^t G^{(m)}(x) \; ,
 \label{Gktdef}
 \\
 G^{(m)}(x) &:=& \sum_{k=0}^{E} x^k \efunc^{(m)}(k)
 \label{Gkmdef}\, .
\eea
The solution is found to consist of simple combinations of the
Hypergeometric function $F(a,b;c;x)$ \cite{EP07b}
\bea
 G^{(m)}(x) &=&  (1-x)^m F(a+m,b+m;c;x)
 \\
 &=& (1-x)^m
 \sum_{l=0}^{E-m}\frac{\Gamma(a+m+l)\Gamma(b+m+l)\Gamma(c)}{\Gamma(a+m)\Gamma(b+m) \Gamma(c+l) (l!)} x^l
 \label{Gmresult}
\eea
with corresponding eigenvalues,
\beq
\lambda_{m} = 1 - m \frac{p_r}{E} -m(m-1) \frac{(1-p_r)}{E^2} ,
 \qquad E \geq m \geq 0 \; .
 \label{eq:evalues}
\eeq
The eigenvalues satisfy $\lambda_{m} > \lambda_{m+1}$ except for
$p_r=0$ when $\lambda_0=\lambda_1=1$.

\begin{figure}[htb!]
\sidecaption
\centering\includegraphics[width=8cm]{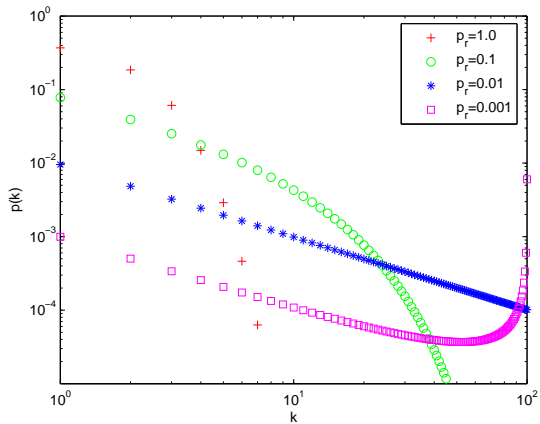}
\caption{Plots of the degree probability distribution function
$p(k)=n(k)/N$ for $N=E=100$ and various $p_r=1$ (red crosses),
$10/E$ (green circles), $1/E$ (blue stars) and $0.1/E$ (magenta
squares). Note that $p_r=1/E$ is almost a pure power law for all
values of $k$.}
 \label{feqDD}
\end{figure}

There is a unique long time equilibrium distribution which can be of
one of two phases, as Fig. \ref{feqDD} shows. For $p_r \lesssim
E^{-1}$ we get a condensate, most individuals attach to a single
artifact. For $p_r \gg E^{-1}$ we get a power law degree
distribution of unit slope, with an exponential cutoff $n(k)\approx
k^{-1} e^{-\zeta k}$ ($\zeta = -\ln \left( 1-p_r \right)$).  There
is a smooth transition between the two except in the $E \ra \infty$
limit.

\begin{figure}[htb!]
\centering
\includegraphics[width=5.5cm]{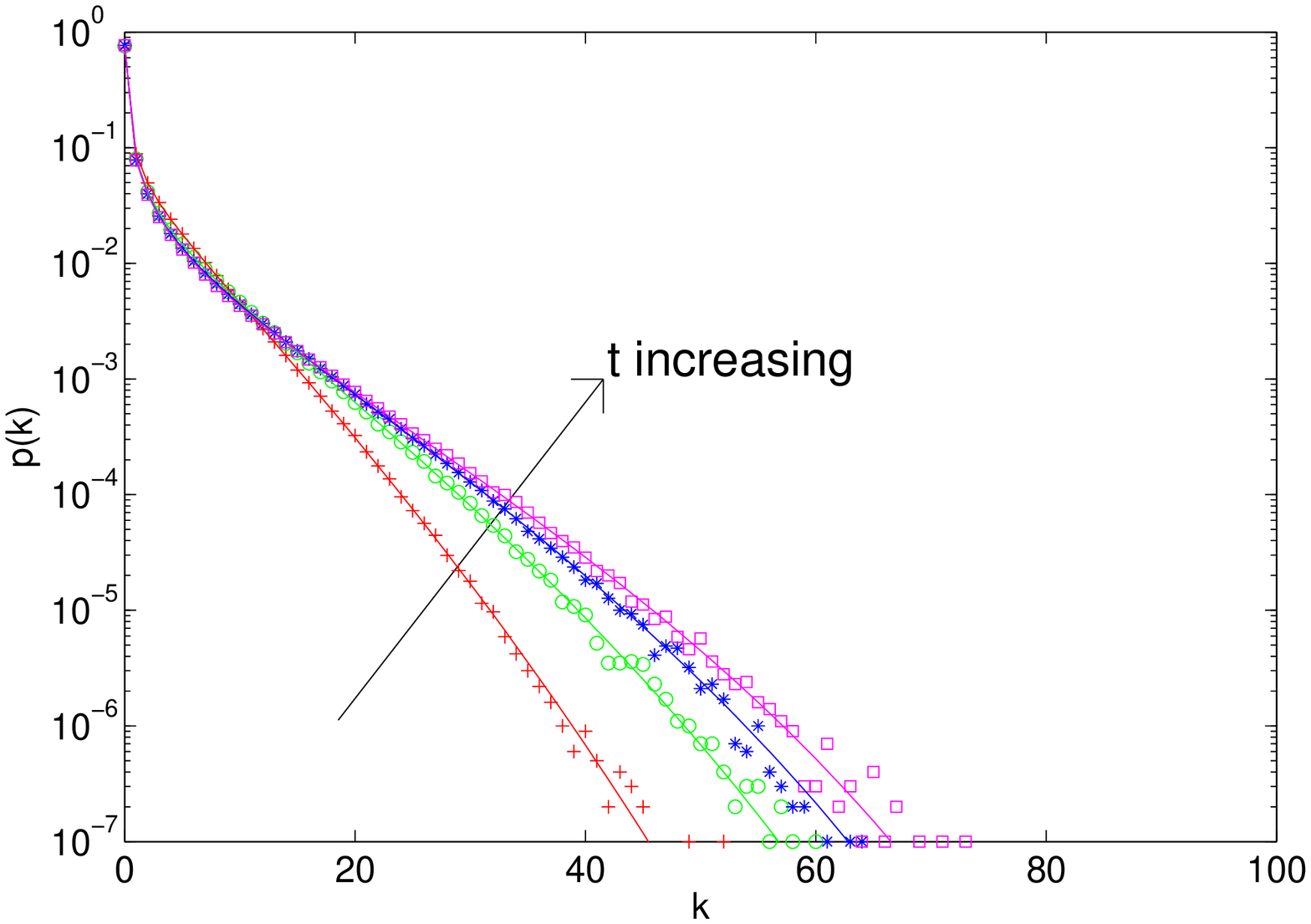}
\includegraphics[width=5.5cm]{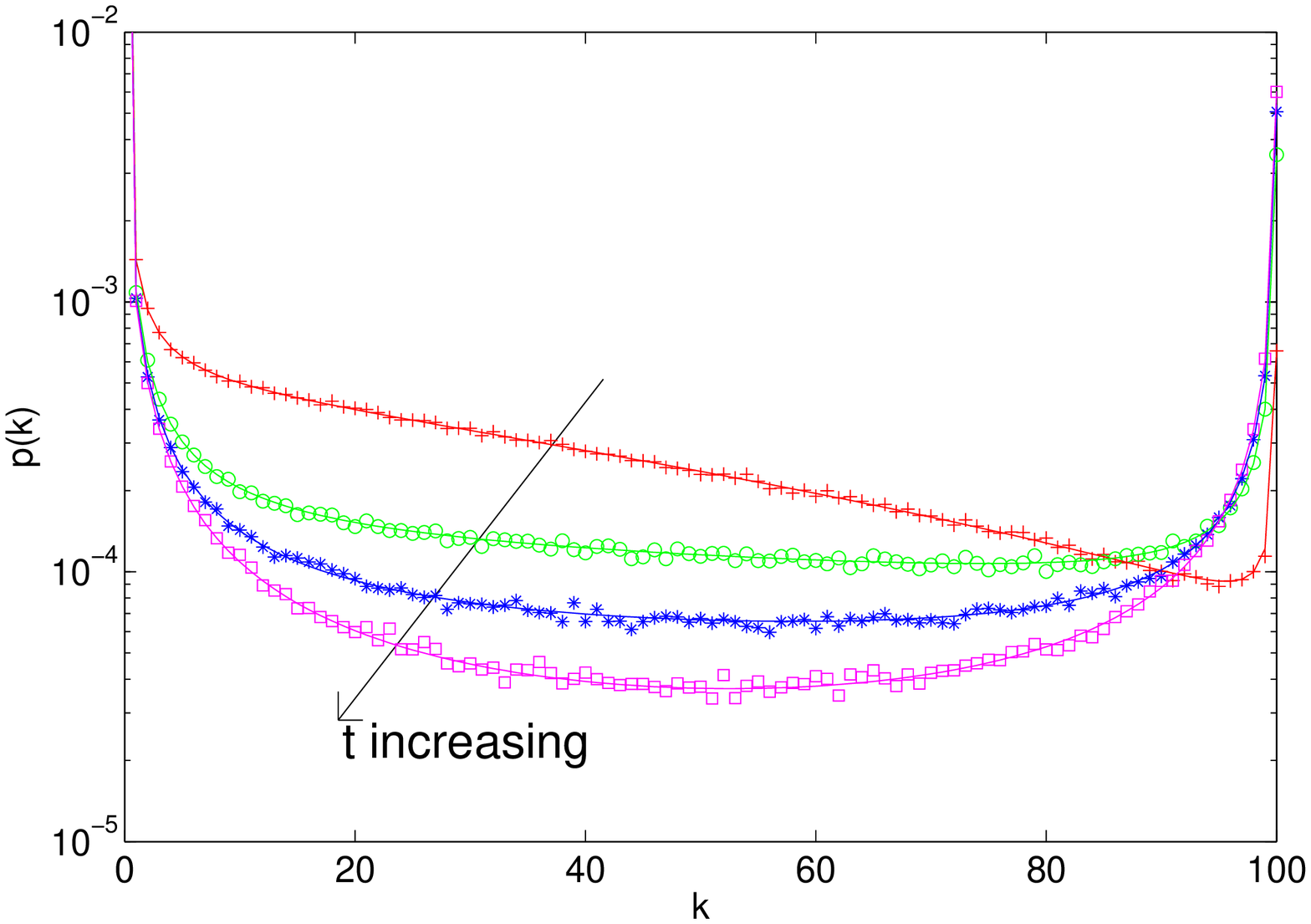}
\caption{Plots of $p(k)$ from simulations (data points) and the
exact analytic results (lines) for $E=N=100$ with $p_r=10/E$ on the
left and $p_r=0.1/E$ on the right.  The results are shown at four
different times: $t\approx\tau_2$ (red, crosses), $t\approx2\tau_2$
(green, circles), $t\approx3\tau_2$ (blue, stars) and to equilibrium
(magenta, squares). The initial configuration has one edge per
artifact. The data points are averages over $10^5$ runs while the
lines are the exact analytic results.} \label{fhalflife}
\end{figure}

Using the fact that the number of artifacts $N$ and the number of
edges $E$ are constant gives $c_0=N$ and $c_1=0$ so that the
eigenmode numbered one ($m=1$) never contributes. Thus the approach
to equilibrium of most quantities occurs on a timescale
\beq
 \tau_2 = - [\ln(\lambda_2)]^{-1}
 \label{tau2def}
\eeq
as illustrated in Fig. \ref{fhalflife}. This means that if we have
rewired most of the edges once and almost never used random
attachment, i.e.\ $p_r \lesssim E^{-1}$, then the approach to
equilibrium is slow, $\tau_2 = O(E^2)$. However for other cases,
$p_r \gg E^{-1}$, the small amount of randomness gives a rapid
approach to equilibrium after every edge has been rewired just a few
times. The initial conditions determine the remaining $c_m$ ($m>1$).

Of particular interest are the homogeneity measures $F_n(t)$ which
are the probability that $n$ randomly chosen but distinct edges all
share the same artifact. These are given by
\beq
 F_n(t) :=   \frac{\Gamma(E+1-n)}{\Gamma(E+1)}
 \left. \frac{d^nG(x,t)}{dx^n}\right|_{x=1} =
  \sum_{k=0}^E \frac{k}{E}\frac{(k-1)}{(E-1)} \ldots
 \frac{(k-n+1)}{(E-n+1)} n(k,t)
 \; .
 \label{Fndef}
\eeq
The properties of the Hypergeometric function mean we can express
the solutions in terms of fixed fractions of a large number of fixed
Gamma functions with all the dependence on time and the initial
conditions is carried by factors of $c_m(\lambda_m)^t$.  Also the
$n$-th homogeneity measure $F_n(t)$ has contributions only from the
eigenfunctions $m\leq n$.

For instance the most useful homogeneity measure is $F_2(t)$:
\bea
 F_2(t) &=&
 F_2(\infty) + (\lambda_2)^t\left( F_2(0) - F_2(\infty) \right) \, ,
 \;\;
F_2(\infty)
 =
 \frac{1+p_r(\langle k\rangle-1)}{1+p_r (E-1)} \, ,
 \label{eqF2tres}
\eea
while the initial conditions set $F_2(0)$.

% *******************************************************
\section{Phase Transitions of Unipartite Graphs in Real Time}\label{sspttime}

\begin{figure}[htb!]
\sidecaption
\centering\includegraphics[width=5cm]{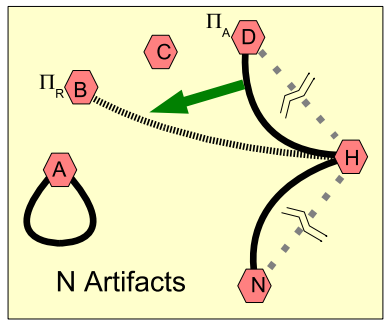}
\caption{One projection of the bipartite graph of Fig.
\ref{fCopyModel4} onto this undirected unipartite graph. Let $a(i)$
be the artifact vertex connected to the individual vertex $i$ in the
bipartite graph. Then we take pairs of individual vertices $(2i)$
and $(2i-1)$ in the bipartite graph and connect their associated
artifacts $a(2i-1)$ and $a(2i)$ in the undirected graph. The
rewiring event of Fig. \ref{fCopyModel4} now become a rewiring of
the (D,H) edge to a (B,H) edge.}
 \label{fCopyModel4uni}
\end{figure}

The construction of Molloy and Reed \cite{MR95} gives a unipartite
graph of a given degree distribution but is otherwise random. In
terms of our bipartite graph this is equivalent to taking pairs of
individual vertices and merging the edge ends (`stubs') coming out
of these individual vertices.  The individual vertices are then
thrown away.  This is illustrated in Fig.\ \ref{fCopyModel4uni}.

The rewiring of our bipartite model is then equivalent to a rewiring
of the projected unipartite graph with the same linear attachment
and removal probabilities, also illustrated in Fig.\
\ref{fCopyModel4uni}.
Since the degree distribution of our artifact vertices is also the
degree distribution of the unipartite graph, all our results can be
applied directly to such graphs.  For instance for $p_r=1$ we
capture the degree distribution of the original Watts and Stogatz
model\footnote{Strictly speaking we choose a random edge to rewire
while Watts and Stogatz \cite{WS98} rewired in a systematic manner.}
\cite{WS98}.

Analytic expressions for the global properties of such random graphs
in the infinite $N$ limit depend on a ratio, $z$, of the second and
first moments of the degree distribution
\cite{NSW01,DMS03a,FFH05,MR95}
\beq
 z(t) := \frac{\ksqav   }{\kav} -1 = (E-1) F_2(t) \, .
  \label{zdef}
\eeq
There is a phase transition in the properties of such infinite
random graphs at $z=1$.  This occurs when there is one tadpole (an
edge connected at both ends to the same vertex) in the unipartite
graph. In particular for $z>1$ the average distance between two
vertices in the giant component, $\langle l \rangle$, may be
estimated to be\footnote{This formula must be adapted from
\cite{FFH05} to take account of the existence of vertices of zero
degree. \tpre{Analytic derivations of such global properties use an
ensemble of graphs over which there is always a finite probability
of getting from any one vertex of degree $k_i>0$ to a vertex of
degree $k_j>0$ in a finite number of steps. In any one graph this
need not be true. On the other hand numerically we measure the
average distance in the largest component of \emph{one} graph
considering only vertices in the largest component.  We then average
this result over the ensemble of graphs.  Numerical evidence
suggests that this numerical measurement has the same qualitative
behaviour as the analytic formula.}} \cite{FFH05}
\beq
 \langle l \rangle = \frac{-2 \langle \ln(k) \rangle + \ln(E)  -
 \gamma_E}{\ln(z)} + \frac{3}{2}
 \, , \qquad \gamma_E \approx 0.5772 \;\; .
 \label{LCdist}
\eeq

For simplicity we consider graphs where $N=E$ which start with each
artifact  connected to only one individual so $n(k,t=0) =
E\delta_{k,1}$.  The projected unipartite graph has $\kav=1$ and
initially $F_2(0)=z(0)=0$.  If $p_r \gg O(E^{-1})$ then the
equilibrium configuration is reached quickly in $t \sim O(\tau_2 ) =
O(E)$ steps.  Only when we start to get a high degree node, so a
condensate is forming and $p_r \lesssim O(E^{-1})$, do we get a
slower approach to equilibrium on a time scale $\tau_2 = O(E^2)$.
The phase transition in infinite random graphs occurs at $z=1$.  In
our case, our projected graphs start from $z(0)=0$ but they reach
$z=1$ very quickly at $t_1 \approx E/2$ unless $(1-p_r) \gg
O(E^{-1})$. That is even if the evolution to the equilibrium
distribution is slow, provided a reasonably large degree node
exists, i.e.\ there is significant amount of copying, a large
component emerges in the projected unipartite graph quickly,
typically at $t_1 \approx E/2$, since
\bea
 t_1 &=& \frac{\ln(1-((E-1)F_{2}(\infty))^{-1})}{\ln(\lambda_2)} \,
 ,
 \\
 &\approx&
   \frac{E}{2(1+p_r(\kav-1))} .\frac{E}{E-1}\, ,
   \qquad (1-p_r) \gg \frac{1}{E} \, .
\eea

\begin{figure}[hbt!]
\sidecaption
\centering
\includegraphics[width=8cm]{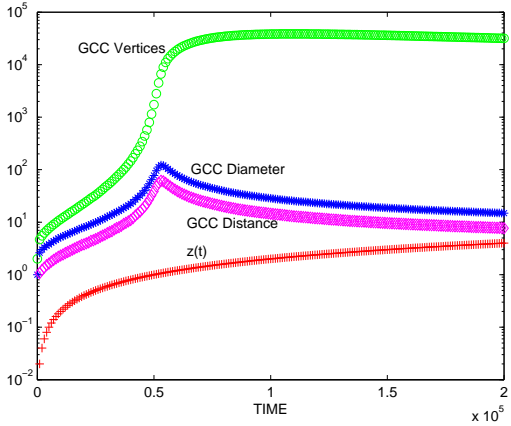}
\caption{Properties of the undirected random graph formed using a
Molloy-Reed type projection \cite{MR95}.  The underlying bipartite
graph has $N=E=10^5$ starting from $F_2(0)=0$ and rewired with pure
copying ($p_r=0.0$). Results are calculated for each instance and
then averaged over a total of 1000 runs.}.
 \label{fPTni1e5}
\end{figure}

The numerical results for the evolution of the properties of the
projected unipartite graph are shown in Fig.\ \ref{fPTni1e5}. The
parameter $z$ reaches the value $1$ at $(t_1/E) \approx 0.5 \pm
0.0002 $ as expected.  This is close to, but not exactly equal to,
$(t_p/E)=0.535 \pm 0.005$, where $t_p$ is the time at which the
average distance and diameter of the largest component peak. The
second derivative in time of the number of vertices in the largest
component also suddenly switches sign at exactly the same time
$t_p$. The value of $z$ at this time is $z(t_p)=1.06 \pm 0.01$.

\begin{figure}[hbt]
\sidecaption
\centering\includegraphics[width=8cm]{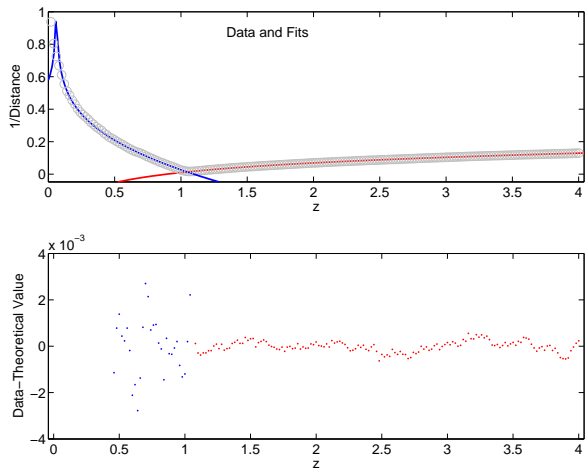}
\caption{The inverse distance of the largest component for the same
projected networks as the previous figure.  The points are the data
(errors are smaller than the symbol size) and the lines are the best
fits to the form $a\ln(z-0.06) + b + cz +dz^2$.  The lower figure
shows the residuals illustrating the good fit.}
 \label{fPTLCinvfit}
\end{figure}

Motivated by the approximate expression for the distance in the
largest component of a large random graph \tref{LCdist}, we find
that the inverse distance for the parameters used in Fig.
\ref{fPTni1e5} is well fitted by the form $a\ln(z-0.06) + b + cz
+dz^2$ but with different values either side of the peak
time\footnote{For early times, $z < z(t_p)$, we have $a=-0.107 \pm
0.006$, $b=0.30 \pm 0.02$, $c=-0.42 \pm 0.04$ and $d=0.14 \pm 0.03$
(fit excluded the four points with lowest $z$ values) while for late
times and $z > z(t_p)$, we have $a=+0.85 \pm 0.02$, $b=0.019 \pm
0.02$, $c=-0.002 \pm 0.06$ and $d=0.0008 \pm 0.0003$. These fits
have $R^2=0.9995$ and $R^2=0.9999$ respectively and errors are at
95\% confidence level. However a polynomial works almost as well, at
least near $t=t_p$.}. The fit is shown in Fig. \ref{fPTLCinvfit}.
The deviations from the predicted $z=1$ transition point seem to be
finite size effects. The peaks are sharper and closer to $z=1$ as
the network get larger\footnote{For $N=E$ and $p_r=0.0$, we find:
$N=10^3$, $t_p/E = 0.66 \pm 0.04$; $N=10^4$,  $t_p/E = 0.57 \pm
0.01$; $N=10^5$, $t_p/E = 0.535 \pm 0.005$. Estimated from an
ensemble of 1000 independent runs for each value of $N=E$.} but with
the same $\kav$, $p_r$ and $F_2(0)=0$.

The network shown is evolved with pure copying $p_r=0$ so in this
case the equilibrium distribution, a complete condensate $F_2 = 1$,
will emerge only on a long time scale of $\tau_2=-\ln(1-2E^{-2})
\sim O(E^2)$.

One way to look at this transition is to use the interpretation of
the model in terms of cultural transmission
\cite{HB03,HBH04,BS05,BLHH07,Evans07,EP07a,EP07b}. In this case the
bipartite graph represents individuals who are choosing artifacts by
either copying the choices made by another individual (preferential
attachment) or by making their own innovation (random attachment).
Suppose we now imagine that each person has two copies of an
artifact.  The unipartite graph is then one expression of the
relationship between objects as defined by the choices made by
individuals.  For instance one could imagine asking people to
categorise their two favourite pairs of shoes and each artifact
could represent a different category, e.g.\ one artifact might
represent black leather lace up shoes.  The unipartite projection
gives a metric in artifact space as defined by the choices made by
the individuals.  The phase transition in the unipartite network
then marks the point where the individuals have reached some sort of
consensus as the artifacts now form a Giant Connected Component
given the metric provided by the individuals' choices.

% -------------------------------------------------------------------
\section{Voter Models and Individual Networks}\label{svoter}

One possible generalisation of our rewiring model is to add a second
graph connecting the individual vertices which we will call the
Individual graph. When an individual rewires using preferential
attachment they copy the artifact chosen by one of their neighbours
in the Individual network.\tnote{I am leaving out discussion of
languages. Should we put it in? *** For example, the Abrams-Strogatz
language model is often studied in the limit when it is equivalent
to a voter model , as used for instance for language evolution
\cite{SCES07}.} With $p_r=0$ and $N=2$ we obtain the basic Voter
model \cite{Liggett99,SR05}. Our model corresponds to having a
complete graph for the individual network but with the addition of a
random rewiring process, $p_r>0$, and an arbitrarily large number of
choices, $N \geq 2$.  Neither of these cases is studied in the Voter
model literature where the focus is on different types of individual
networks and any analytic results are only available for the $E \ra
\infty$ limit \cite{SR05}\footnote{For $p_r>0$ we can think of our
model as including two graphs. The first, as mentioned above, is a
graph connecting Individuals. Preferential rewiring is done by
performing a random walk of length one on this graph, and copying
the choice of the resultant Individual. The second graph connects
the Artifacts \cite{EP07b}.  If this graph is a complete graph (with
tadpoles) then a random walk on the graph gives the random
attachment $p_r$ term appearing in $\Pi_A$ of (\ref{PiRPiAsimple}).
One may imagine many practical problems where the Artifact network
is not so trivial. For this paper, however, we only consider the
case of a complete Artifact graph.}

\begin{figure}[hbt!]
 \sidecaption
 \centering
 \includegraphics[width=6.5cm]{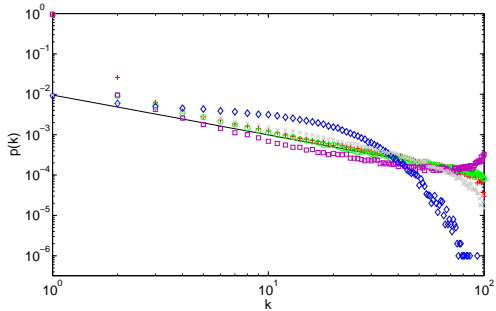}
\caption{Equilibrium artifact degree distribution $p(k)$ for
different Individual graphs of 100 vertices and average degree 4:
Erd\H{o}s-R\'{e}yni (red pluses), Exponential (green circles),
Barab\'asi-Albert (purple squares), periodic lattices of two (grey
crosses) and one (blue diamonds) dimension. The line is the analytic
result for a complete Individual graph while the other results are
taken over an ensemble of $10^4$ Individual graphs. $N=E=100$,
$p_r=1/E$.}
 \label{fvotereq}
\end{figure}

Results for the equilibrium distribution show that it is
qualitatively unchanged by the type of individual graph\footnote{In
this article, the lattices are cubic ($\mathbb{Z}^d$).  In Fig.\
\ref{fvotereq} next-to-nearest and nearest neighbours are connected
in the one-dimensional ring.  In all other cases the only nearest
neighbours connected in the lattice Individual graphs. The
Exponential and Barab\'asi-Albert graphs are connected graphs
Individual graph degree distributions of $p_\mathrm{ind}(k) \propto
\exp\{-\zeta k\}$ and $p_\mathrm{ind}(k) \propto [k(k+1)(k+2)]^{-1}$
respectively. The results also support the claim in
\cite{Evans07,EP07a,EP07b} that some results for a Minority game
played on an Erd\H{o}s-R\'{e}yni graph can be understood in terms of
our results for the rewiring model.} except for the case of a one
dimensional ring, as shown in Fig.\ref{fvotereq}.

We will use two quantities to study the behaviour of the model. Our
quantity $F_2$ of \tref{Fndef} is a measure of the global
homogeneity. An equivalent measure which takes account of the local
properties of the Individual network is the average interface
density, $\rhoexp$, the probability that any two individual vertices
connected by the Individual graph have a different artifact. If the
graph is complete or if the Individual graph is ignored ($p_r=1$)
then from \tref{eqF2tres} we have $\rhoexp_t=1-F_2(t)$.\tnote{We
would also expect this to be true for ensembles of random graphs
whatever their degree distribution. Is this true for random graphs?
Or is it a length dependent statement? Try randomising some lattices
etc.  To what extent our analytic approximations deviate from the
true $\rhoexp$ will be the focus of this section.} Otherwise for two
reasons we expect that $\rhoexp_t \neq 1-F_2(t)$  and that both
would both differ from value obtained for a complete Individual
graph as derived from \tref{eqF2tres}. First because of the explicit
reference to the Individual graph in the definition of $\rhoexp_t$
but not in $F_2$\tnote{Are the $\rhoexp$ analogues of higher $F$'s
interesting? i.e. the average local homogeneity measures.  For
instance the probability that two neighbours both share the same
artifact as the central Individual.}. Second, the structure imposed
by the Individual graph will, in general, effect both the evolution
timescale and, for $p_r>0$, equilibrium degree distributions as
compared to the complete Individual graph case.

\begin{figure}[hbt!]
 \sidecaption
 \centering
\includegraphics[width=8cm]{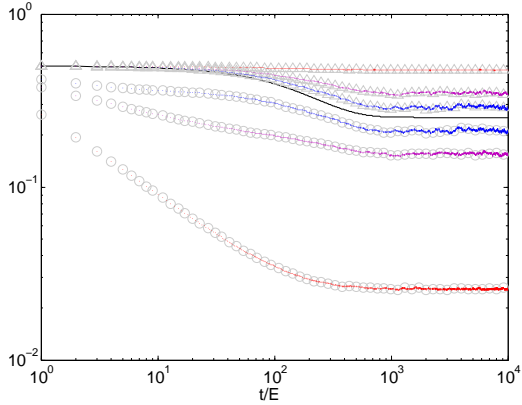}
\caption{Homogeneity measures for various lattices against $t/E$.
The black solid line represents the analytic $ 1-F_2(t)$ for $N=2$,
$p_r=1/E$ and $E=729$. Numerical results for $1-F_2(t)$ (triangle
highlights) are plotted for 1-d (red), 2-d (purple) and 3-d (blue)
regular lattices. The average interface densities $\rhoexp$ are
plotted as circles. Averaged over $1000$ runs.}
 \label{flatcomp}
\end{figure}

We can see these differences if we compare the equilibrium values
reached on lattices of different dimensions but with some randomness
present (otherwise $\rhoexp_t= (1-F_2(t))$ because both are zero).
As Fig.\ref{flatcomp} and table \ref{tlatcomp} show, the local and
global homogeneity measures $\rhoexp$ and $(1-F_2)$ are close to the
analytic result for large dimension lattices with short network
distances.  As we take lattices of smaller dimension, $F_2$ gets
much larger than the analytic result, and $\rhoexp$ much smaller.
Table \ref{tprcomp} shows a similar effect as we increase $p_r$.

\begin{table}
\centering
\begin{tabular}{r|c|c|c|c}
Dim & $t_0(F_2)/\tau_2$ &  $t_0(\rho)/\tau_2$ & $1-F_2(\infty)$ &
$\rho(\infty)$  \\ \hline
 1d & 1.25 (3)          & 0.0241 (4)          & 0.47466 (2)     & 0.0261  (1) \\
 2d & 1.19 (1)          & 0.74 (1)            & 0.3494  (1)     & 0.1558  (1) \\
 3d & 1.10 (1)          & 1.06 (1)            & 0.2898  (1)     & 0.2120  (1)
\end{tabular}
\caption{Table of time scales and the limiting value of the
evolution of $1-F_2$ and $\rho$ where the Individual graphs are
periodic lattices with nearest neighbour connections only. The
complete graph has $1-F_2(t=\infty) \approx 0.25017$ and $\tau_2
\approx 1.3295\time 10^5$. Extracted from the data of Fig.
\ref{flatcomp} by fitting to $a \exp(-t/t_0)+c$  with the estimated
error in the last digit give by the numbers in brackets.}
 \label{tlatcomp}
\end{table}

The time scale of the approach to consensus is often studied in
Voter models.  If $p_r>0$, so there is no absolute consensus, the
approach to equilibrium, as measured by $F_2$ and $\rho$,  is
controlled solely by the time scale of the second eigenvalue
$\tau_2$ of \tref{tau2def} if the Individuals are connected by a
complete network.  For general Individual networks the evolution of
the global $F_2$ or local $\rhoexp$ takes the same form  as
$\tref{eqF2tres}$, $a \exp(-t/t_0)+c$. However, as one might expect,
the formation of small patches of consensus between nearest
neighbours, measured by $\rhoexp$, happens faster than the emergence
of a global consensus, as measured by $F_2$.  This is accentuated if
there is a large distance between individuals\tnote{Compare with EXP
or BA graphs of K=2. Try random graphs of fixed degree} as the
comparison between lattices of different dimensions in
Fig.\ref{flatcomp} and in table \ref{tlatcomp} show.  Varying $p_r$
also shows that local equilibration is faster than global but there
does seem to be a marked difference between $p_r \ll 1/E$ and $p_r
\gtrsim 1/E$ as table \ref{tprcomp} shows.\tprenote{See also Fig.\
\ref{fprcomp}.} For $p_r E \ll 1$, local equilibration is a little
slower than occurs on complete graph. However for $p_r E \gtrsim 1$,
this randomness brings local equilibrium an order of magnitude
faster than was the case with a complete graph. It shows that a
little bit of randomness can speed up local equilibration but not if
an overwhelming consensus is going to emerge.

\begin{table}
\centering
\begin{tabular}{r||c|c|c||c|c|c}
$p_r/E$ &  $\tau_2$ & $t_0(F_2)/\tau_2$ & $t_0(\rho)/\tau_2$
                    &  $1-F_2(\infty)$ exact & $1-F_2(\infty)$   & $\rho(\infty)$ \\
\hline
0   & 79999 &    2.131  (6) &  1.8   (1) & 0        & 0.0003 (3)   & 0.003  (3) \\
0.1 & 72743 &    1.988  (5) &  1.81  (2) & 0.04546  & 0.0827 (3)   & 0.0403 (4) \\
1   & 40050 &    1.34   (3) &  0.15  (2) & 0.25031  & 0.3375 (3)   & 0.167  (1) \\
10  &  7289 &    1.01   (6) &  0.145 (3) & 0.45558  & 0.47433 (4) & 0.2634 (1) \\
100 &   794 &    0.7    (2) &  0.40  (1) & 0.49628  & 0.49712 (1) & 0.3813 (1)  \\
\end{tabular}
\caption{Table of time scales in units of $\tau_2$ and limiting
value in units of the exact value for $c=F_2(t=\infty)$ for 400
individuals connected by a square lattice. Data averaged over $5000$
runs for $p_r=0$ and $p_r=1/E$, and $1000$ runs for all others. It
was fitted to $a \exp(-t/t_0)+c$.} \label{tprcomp}
\end{table}

\begin{table}
\centering
\begin{tabular}{c|cc||c|c|c|c}
&      &         & \multicolumn{2}{c|}{$N=2$} & \multicolumn{2}{c}{$N=10$} \\
Dim &   $E$& $\tau_2$& $t_0(1-F_2)/\tau_2$ & $t_0(\rho)/\tau_2$ & $t_0(1-F_2)/\tau_2$ & $t_0(\rho)/\tau_2$ \\
   \hline
   &    100 &   4999.5  &  18.6 (1) & 0.42  (2) &   18.9 (1)  & 0.41  (1) \\
1d &    200 &  19999    &  37.9     & 0.21  (1) &   39.3      & 0.21  (1) \\
   &    400 &  79999    &  76.3     & 0.18  (1) &   75.1      & 0.18  (1) \\
   &   1000 & 500000    & 115.7     & 0.069 (2) &  137.8      & 0.070 (3) \\
   \hline
   &  400 & 79999   & 2.131 (3) &    1.8 (1) & 2.109 (2) &  1.8 (1) \\
2d &  900 & 405000  & 2.3       &    2.1     & 2.3       &  2.1     \\
   & 2500 & 3130000 & 2.8       &    2.1     & 2.8       &  2.1
\end{tabular}
\caption{Table of time scales in units of $\tau_2$ for $E$
individuals connected by a one- or two-dimensional torus found by
fitting the data to $a \exp(-t/t_0)+c$.  For two and ten types of
artifact, $p_r=0$.  Data was averaged over $1000$ runs for the
largest lattices down to $50$ runs for the smallest lattices. Where
the error is known reliably, the numbers in brackets specify the
error in the last digit.} \label{tEnacomp}
\end{table}

In table \ref{tEnacomp} we see that the time scales for the
exponential decay obtained from fitting our data are roughly in line
for the predictions made for the completion time in this model
\cite{Liggett99,SR05,Krap92} on a lattice, $t_0 \sim O(E)$ for a
one-dimensional lattice, $t_0 \sim O(\ln(E))$ in two dimensions.
However some discrepancies suggest more work is needed.

\begin{figure}[hbt!]
 \sidecaption
 \centering
\includegraphics[width=7cm]{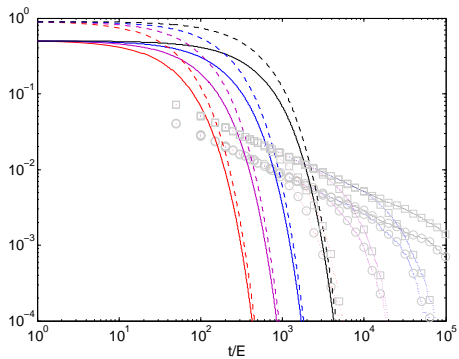}
\caption{Plots of the analytic $(1-F_2(t))$ (lines) and $\rhoexp_t$
for 1-d periodic lattices against $t/E$ for $p_r=0$ and $E=100$
(red, far left), $200$ (purple), $400$ (blue) and $1000$ (black, far
right). The lower circles and the solid lines represent $N=2$ while
the higher squares and dashed lines are for $N=10$. Data are
averages over $1000$, $1000$, $500$ and $100$ runs for increasing
lattice sizes respectively.}
 \label{f1dna}
\end{figure}

The main result to draw from table \ref{tEnacomp} is that the
evolution towards equilibrium, its time scale and final value, are
independent of the number of artifacts.  This is to be expected at
small $p_r$ given our linear attachment probabilities as this gives
our model certain scaling properties \cite{EP07b}.  Suppose we have
the consensus emerging picking out one of our $N$ artifacts and we
merge the remaining $N-1$ artifacts into one artifact.  The
probability of an individual copying the consensus artifact or one
of the remaining artifacts is exactly the same as if we had a model
with $N=2$ and the same $(1-p_r)$.  The only difference is that when
a random innovation event occurs, with probability $p_r$, the
non-consensus artifacts are preferred to the single consensus
artifact by a factor of $(N-1)$.  Thus for $N \gg 2$ the random
events are more likely to destroy the emerging consensus than in the
Voter model but \emph{only} if $p_r \gg 0$.  For the extreme case of
$p_r=0$ we see the expected lack of dependence on $N$ in table
\ref{tEnacomp}. The only effect of increasing the number of
artifacts in our results comes from starting from a homogeneous
initial condition so $F_2 = 1/N$ which is further away from $F_2=1$
and consensus, see Fig.\ \ref{f1dna}.

% -------------------------------------------------------------------
\section{Two Types of Individual}\label{stwoind}

Another variation of our original model is to introduce two types of
individual, labelled $X$ and $Y$.  At each time step we first pick
which type of individual to update; with probability $q_x$ we select
at random one the $X$-type individuals.  We rewire its artifact end,
choosing its arrival artifact in one of three ways: at random, by
copying the existing choice of one its own type of individual, or
finally copying the existing choice made by a random individual of
the opposite type.  These arrival probabilities may be different for
the two types so we have four independent arrival probabilities and
one departure probability.  Add in the freedom to choose different
numbers of $X$ and $Y$ individuals, $E_x$ and $E_y$, and $N$ the
number of artifacts, we find we have eight free parameters.  The
degree distribution is now $n(k_x,k_y;t)$, the number of artifact
vertices which have $k_x$ ($k_y$) edges to $X$ ($Y$) type vertices
at time $t$.

The question is can we solve this system analytically?  By keeping
our probabilities linear in degree and because our normalisations
are constants of the evolution, our mean field equation is again
exact, for the same reasons as in the original model \cite{EP07b}.
Writing in terms of the generating function $G(x,y,t) :=
\sum_{k_x=0}^{E_x} \sum_{k_y=0}^{E_y} x^{k_x} y^{k_y} n(k_x,k_y,t)$
we find that we can again split this into $(E_x+1)(E_y+1)$
eigenfunctions which we label with a pair of indices $(M,A)$:
\beq
 \GMa (x,y) := \sum_{i=0}^{E_x} \sum_{j=0}^{E_y} (x-1)^i (y-1)^j
 \fMa_{ij} \, ,
\eeq
where $\fMa_{ij}$ are constants. The eigenfunctions satisfy a
two-dimensional second order PDE. We have not found a full solution
but we can reduce this to a one-dimensional problem.  Since we
express our eigenfunctions in powers of $(x-1)$ and $(y-1)$, the
eigenfunctions satisfy\footnote{We use this to define our label
$M$.} $\fMa_{ij}=0$ if $i+j <M$ for any integer $0 \leq M \leq
E_x+E_y$.  At the same time the equations for the coefficients
$\fMa_{ij}$ and eigenvalues $\lMa$, where $i+j=M$, involves no
coefficients where $i+j>M$. Thus the label $A$ indexes the allowed
values of $i$ and $j$ given the constraint $i+j =M$. Finding the
eigenvalues is therefore a matter of solving a set of
$(\min(E_x,E_y)+1)$ linear equations. This also gives the
coefficients of the eigenfunctions for $i+j=M$. The remaining values
$i+j>M$ may be found iteratively though this is generally a
non-trivial problem.

However, we have seen that much information is encoded by the first and second moments of the degree distribution.  The general
homogeneity measures are given by
\beq
 F_{mn}(t) :=
 \frac{\Gamma(E_x+1)}{\Gamma(E_x-m+1)}
 \frac{\Gamma(E_y+1)}{\Gamma(E_y-n+1)}
 \left. \frac{\partial^{m+n} G(x,y;t)}{\partial x^m \partial y^n}\right|_{x=y=1}
\eeq
So we only need the three second order homogeneity measures,
$m+n=2$.  These have contributions only from the $i+j\leq 2$
coefficients and therefore only the $M \leq 2$ eigenfunctions
contribute.  The system of equations for such coefficients reduces
to solving for the eigenvalues and eigenfunctions of a three
dimensional system, which has an exact, if lengthy, algebraic
solution. The basic results though are that the only equilibrium
solution is given by the single $M=0$ eigenfunction where
$\lzero=1$, $\fzero_{00}=N$, $\fzero_{10}=E_x$, $\fzero_{01}=E_y$,
and the three coefficients $\fzero_{20}, \fzero_{11},\fzero_{02}$
satisfy
\bea
  \fvec^{(0)}
  &=&
  \begin{pmatrix} \fzero_{20} \\ \fzero_{11} \\ \fzero_{01} \end{pmatrix}
  =
  \Tmat^{-1} \begin{pmatrix}
  \alpha_x(1+a_x)(E_x-1)  \\
  \alpha_x(d_x - a_x/E_x)E_y + \alpha_y(d_y - a_y/E_y)E_x   \\
  \alpha_y(1+a_y)(E_y-1)
 \end{pmatrix} \, ,
 \\
  \Tmat &:=&
 \begin{pmatrix}
 -2 \alpha_x (1+E_x^{-1}) & - \alpha_x d_x (1 -E_x) & 0 \\
 - 2 \alpha_y d_y & \alpha_x\beta_x+\alpha_y\beta_y &  - 2 \alpha_x d_x \\
 0 &  - \alpha_y d_y (1 -E_y) &  -2 \alpha_y (1+E_y)  \\
 \end{pmatrix}
 \, ,
\\
a_x
 &=&
 \frac{p_{r_x}}{p_{p_{xx}}}\langle k_x \rangle
 \, , \;
 d_x = -\frac{p_{p_{xy}}}{p_{p_{xx}}}\frac{E_x^2}{E_y}
 \, , \;
%\nonumber \\
 \alpha_x  =
  \frac{q_x p_{pxx}}{(E_x)^2}
  \, , \;
 \beta_x = \frac{E_x (1- p_{pxx})}{p_{pxx}}  .
 \label{coeff}
\end{eqnarray}
Switching labels $(x \leftrightarrow y)$ gives the similar $y$
subscript parameters.

The $M=1$ eigenfunctions again give no contribution to any physical
quantity since the first moments are constant. The second moments
are given in terms of the lowest coefficients of one of three $M=2$
eigenfunctions, the $\ftwoa_{20}, \ftwoa_{11},\ftwoa_{02}$ (assuming
$E_x,E_y>1$), which satisfy
\bea
 (\ltwoa-1)\fvec^{(2A)} & = & \Tmat \fvec^{(2A)}
 \, , \qquad
 \fvec^{(MA)} = \begin{pmatrix} \fMa_{20} \\ \fMa_{11} \\ \fMa_{01}
\end{pmatrix}
\, .
\eea

There is a large parameter space to investigate but there are a few
obvious limits.  First one can scale the probabilities in proportion
to the number of edges of each type so $q_a= E_a/E$, $p_{ra} = p_r$,
$p_{pab}=(1-p_r)E_b/E$ where $a,b \in \{ x,y\}$, $E=E_x+E_y$.  One
can see then that the total degree distribution given by $G(x,x)$ is
exactly as we had in the single type model.  However we can now
investigate the `chemical equilibrium' as the distribution of $X$
and $Y$ types, given by derivatives of $G(x,1)$ and $G(1,y)$
respectively, will evolve differently if the initial conditions are
different for each type.  Another simple example is where
$p_{xx}=p_{yy}=0$ which encodes the ``complete bipartite graph
example'' of \cite{SR05}.  Our method allows one to extract exact
expressions for the whole time evolution, not just order of
magnitude estimates for the equilibration time.

% ****************************************************************
\section{Conclusions}\label{scon}

In this paper we have looked at a variety of extensions to the basic
network rewiring model of \cite{Evans07,EP07a,EP07b}. Studying the
projection onto a unipartite graph gives us \emph{exact} expressions
for the time evolution of a finite sized system through a
transition.

We have also shown that adding an Individual network leaves the
qualitative behaviour of the model is unchanged in terms of $F_2$.
However quantitative differences are highlighted by comparison
against the case of a complete graph for which our previous analytic
work \cite{Evans07,EP07a,EP07b} provides exact analytic results.
What we learn from this model is that the consensus (the condensate)
may not be perfect, $1 \gtrsim p_r E
>0$, and it may emerge very slowly $\tau_2 \sim O(E^2)$, but an
effective consensus is always reached very quickly $t_1 \sim O(E)$.

Finally we have shown how some progress can be made on solving
models with more than one type of edge.  In particular we show how
the various homogeneity measures $F_{mn}$ may be found exactly.

% *****************************************************************
% Uses BibTeX
%
%\newpage
\bibliographystyle{unsrt}
\bibliography{rweccs07bib}

\newpage
% *******************************************************
\appendix
\section*{Additional Figures}
These were not included in the proceedings.  Data in some of the
tables was derived from these curves.

\begin{figure}[hbt!]
 \sidecaption
% \centering
\includegraphics[width=8cm]{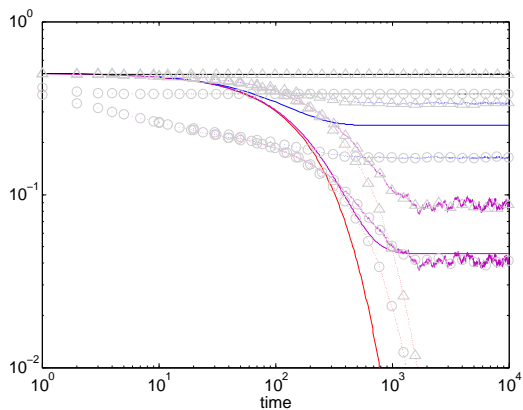}
\caption{Plots of $1-F_2(t)$ (triangles) and $\rhoexp$ (circles) on
a 2-d periodic lattice with side $L=20$ and $N=2$ with attachment
probabilities $p_r=0$ (red), $p_r=0.1/E$ (purple), $p_r=1/E$ (blue)
and $p_r=100/E$ (black). Solid coloured lines are the equivalent
analytic results for $1-F_2(t)$ on a complete graph. Averaged over
$5000$ runs for $p_r=0$ and $p_r=1/E$ and $1000$ runs for all
others. Data used for table \ref{tprcomp}.}
 \label{fprcomp}
\end{figure}

\begin{figure}[hbt!]
 \sidecaption
% \centering
\includegraphics[width=8cm]{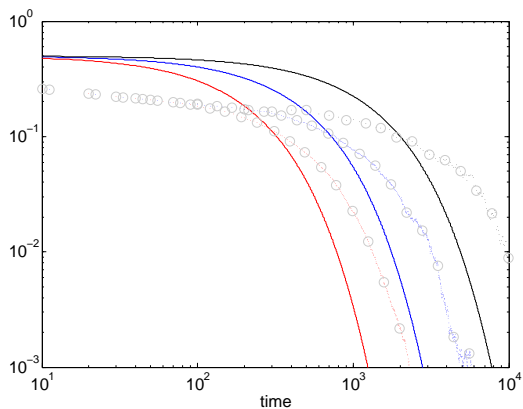}
\caption{Analytic $1-F_2(t)$ (solid lines) for a complete graph
compared against numeric $\rhoexp$ (circles) on a 2-d periodic
lattice with $N=2$ and $p_r=0$. $L=20$ (red), $L=30$ (blue) and
$L=50$ (black). Averaged over $5000$, $500$ and $50$ runs
respectively.  Data used for table \ref{tEnacomp}.}
 \label{fEcompF2}
\end{figure}

\begin{figure}[hbt!]
 \sidecaption
% \centering
\includegraphics[width=8cm]{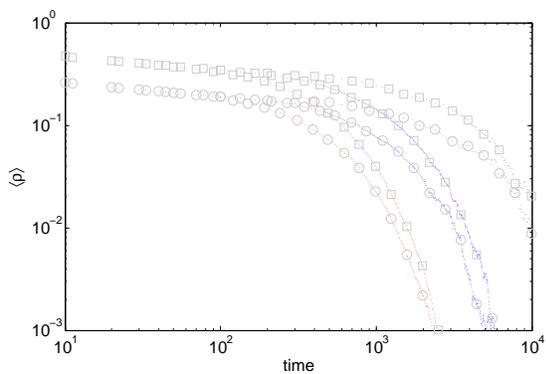}
\caption{Plots of $\rhoexp$ on a 2-d periodic lattice with $p_r=0$
for $N=2$ (circles) and $N=10$ (squares). $L=20$ (red), $L=30$
(blue) and $L=50$ (black). Averaged over $5000$, $500$ and $50$ runs
respectively. Data used for table \ref{tEnacomp}.}
 \label{fE2drho}
\end{figure}

\end{document}